\documentclass[aps,prd,12point,twocolumn,nofootinbib,showpacs,superscriptaddress]{revtex4-1}
\def\theequation{\arabic{section}.\arabic{equation}}
\usepackage{psfrag}
\usepackage{subfigure}
\usepackage{color}
\usepackage{mathrsfs}
\usepackage{graphicx}
\usepackage{amssymb}
\usepackage{amsmath, amsthm}
\usepackage{epstopdf}
\usepackage{hyperref}
\usepackage{enumerate}

\newcommand{\be}{\begin{equation}}
\newcommand{\ee}{\end{equation}}

\begin{document}
\def\theequation{\arabic{section}.\arabic{equation}} 

\title{New inhomogeneous universes in scalar-tensor and $f({\cal 
R})$ gravity}

\author{Valerio Faraoni}
\email[]{vfaraoni@ubishops.ca}
\affiliation{Department of Physics and Astronomy, Bishop's 
University, 
2600 College Street, Sherbrooke, Qu\'ebec, Canada J1M~1Z7
}
\author{Shawn D. Belknap-Keet}
\email[]{sbelknapkeet02@ubishops.ca}
\affiliation{Department of Physics and Astronomy, Bishop's 
University, 2600 College Street, Sherbrooke, Qu\'ebec, 
Canada J1M~1Z7
}



\begin{abstract}
A new family of spherically symmetric 
inhomogeneous solutions of Brans-Dicke gravity is 
generated using the Fonarev solution of general relativity 
as a seed and a map from the Einstein to the Jordan 
conformal frame. The Brans-Dicke scalar 
field self-interacts with a power-law or inverse power-law 
potential in the Jordan frame. This 4-parameter family of 
geometries, which is dynamical and asymptotically 
Friedmann-Lema\^{i}tre-Robertson-Walker, contains as 
special cases two previously known classes of 
solutions and solves 
also the field equations of $f({\cal R})={\cal R}^n$ 
gravity.
\end{abstract}


\maketitle

\section{Introduction} 
\label{sec:1}
\setcounter{equation}{0}

General relativity (GR) has been tested with good precision 
and its prediction of gravitational waves has received a 
spectacular experimental confirmation with the recent {\em 
LIGO} detections \cite{LIGO1, LIGO2}. However, the theory 
is not tested at most spatial and temporal scales and 
curvature regimes \cite{Bertietal2015, Psaltis}. What is 
more, GR does not agree with quantum mechanics and all 
attempts to quantize it produce, in their low-energy limit, 
theories which deviate from GR. The most compelling 
motivation to go beyond Einstein theory, however, comes 
from 
cosmology: within the context of GR the present 
acceleration of the universe discovered with type Ia 
supernovae can only be explained with an enormously 
fine-tuned cosmological constant or with a completely {\em 
ad hoc} dark energy fluid as a matter source in the 
Einstein equations. A viable alternative consists of 
modifying gravity at cosmological scales while leaving 
untouched the predictions of GR at small scales. The most 
popular class of theories achieving this goal is $f({\cal 
R}) $ gravity (where ${\cal R}$ is the Ricci scalar of the 
metric connection) 
\cite{CCT}, see \cite{reviews} for reviews.

The prototype of alternative gravity is Brans-Dicke theory 
\cite{BD}, which has been generalized to the wider class of 
scalar-tensor theories \cite{ST} and contains as a 
fundamental 
variable a gravitational scalar field $\phi$ in addition to the 
metric tensor $g_{ab}$. The wide class of $f({\cal R})$ theories is 
a subclass of scalar-tensor gravity. When attempting to understand 
these theories, spherically symmetric analytic solutions play an 
important role. Alternative theories of gravity which attempt to 
explain the current acceleration of the cosmic expansion without 
dark energy have a built-in time-dependent cosmological 
``constant'' and spherical objects
 in these theories are not isolated, but are asymptotically 
Friedmann-Lema\^itre-Robertson-Walker (FLRW) and are 
dynamical. Even in GR, exact solutions of the field 
equations representing dynamical inhomogeneous universes 
are rare and their physical interpretation is often 
puzzling 
(\cite{Krasinskibook, mylastbook} and references 
therein). Here we take one such solution of GR, the Fonarev 
inhomogeneous universe sourced by a matter scalar field 
with an exponential potential \cite{Fonarev, MaedaFonarev}, 
and we use it as a seed to generate a family of new 
solutions of Brans-Dicke gravity with a power-law (or 
inverse power-law) potential. We then show that this family 
of geometries is also a solution of a class of $f({\cal 
R})$ theories. Extra motivation for this work comes from 
the old idea that the gravitational constants of nature may 
not be constant after all \cite{Dirac, Barrowbook}, and 
scalar-tensor gravity provides an arena in which the 
gravitational coupling strength is dynamical. In this 
context, inhomogeneous universes are useful to probe 
spatial variations of the gravitational ``constant'', which 
was the motivation behind Ref.~\cite{CMB} containing a 
geometry which corresponds to a special case of the new 
family of solutions that we introduce here.

Following the notation of Ref.~\cite{Waldbook} and using 
units in which Newton's constant $G$ and the speed of light 
are unity, the action 
of vacuum Brans-Dicke theory in the Jordan frame is 
\cite{BD}
\be \label{BDaction}
S_\text{BD} = \int d^4x \, \frac{ \sqrt{-g}}{16 \pi}
\left(\phi  \mathcal{R}-\frac{\omega}{\phi} \nabla^c\phi
\nabla_c\phi  -V(\phi)\right) \,,
\ee
where $\phi$ is the Brans-Dicke scalar field (approximately 
equivalent to the inverse of the gravitational coupling 
$G_{\text{eff}}$) with potential $V(\phi)$, $\omega$ is the 
constant Brans-Dicke parameter, and $g$ is the 
determinant of the spacetime 
metric $g_{ab}$. The variation of the action~(\ref{BDaction}) 
generates the Brans-Dicke field equations {\em in vacuo}  
\cite{BD}
\begin{eqnarray}
R_{ab}-\frac{{\cal R}}{2} g_{ab} &=& \frac{\omega}{\phi^2}
\left( \nabla_a\phi \nabla_b\phi -\frac{1}{2}\, g_{ab}
\nabla^c \phi\nabla_c\phi \right)  \nonumber\\
&&\nonumber\\
&\, & +\frac{1}{\phi} \left( \nabla_a\nabla_b\phi -g_{ab} 
\Box \phi 
\right) -\frac{V}{2\phi}\, g_{ab} \,,\label{BDfe} 
\nonumber\\
&&\\
\Box \phi  &=& \frac{1}{2\omega+3} \left( \phi \, 
\frac{dV}{d\phi} 
-2V \right) \,. \label{box}
\end{eqnarray}
(The original Brans-Dicke theory \cite{BD} did not include a 
potential $V$ for the Brans-Dicke field $\phi$.) The more general 
class of scalar-tensor theories \cite{ST} promotes the Brans-Dicke 
parameter $\omega$, which is constant in the original 
Brans-Dicke theory, to a function of the scalar $\phi$.

Another representation of scalar-tensor gravity, the 
Einstein frame \cite{Dicke}, is widely used. By performing the 
conformal 
transformation of the  metric
\be \label{metric transformation}
g_{ab} \rightarrow \tilde{g}_{ab} = \phi \, g_{ab},
\ee
and the scalar field redefinition
\be
\phi \rightarrow \tilde{\phi}=\sqrt{\frac{|2\omega+3|}{16
\pi}} \, \ln \left( \frac{\phi}{\phi_{*}}\right) \,,
\ee
where $\phi_{*}$ is a constant and $\omega\neq -3/2$, the 
Brans-Dicke
action~(\ref{BDaction}) assumes its
Einstein frame form
\be
S_\text{BD} = \int d^4x\sqrt{-\tilde{g}}\left[
\frac{\tilde{\mathcal{R}}}{16 \pi}-\frac{1}{2} \, 
\tilde{g}^{ab}\nabla_a\tilde{\phi} \nabla_b\tilde{\phi}  
-\tilde{V}(\tilde{\phi}) \right]  \,, \label{BDactionEframe}
\ee
where
\be
\tilde{V}(\tilde{\phi}) = \frac{V(\phi)}{\phi^2}\left|_{\phi=\phi( \tilde{\phi}) } 
\right. \,. \label{ppotential} 
\ee
In the following, Einstein frame quantities will be denoted by a 
tilde. The action~(\ref{BDactionEframe}) is formally the 
Einstein-Hilbert action coupled to a matter scalar field which has  
canonical kinetic energy density. The Einstein frame field 
equations {\em 
in vacuo} are
\begin{eqnarray}
\tilde{R}_{ab}-\frac{1}{2} \, \tilde{g}_{ab} \tilde{
{\cal R}} &=& 8\pi \left( \nabla_a \tilde{\phi} \nabla_b
\tilde{\phi} -\frac{1}{2} \, \tilde{g}_{ab}
\tilde{g}^{cd} \nabla_c \tilde{\phi}\nabla_d \tilde{\phi}
\right)  \nonumber\\
&&\nonumber\\
&\, & -\tilde{V}(\tilde{\phi}) \tilde{g}_{ab} \,, \label{Eframefe}
\end{eqnarray}
\be
 \tilde{g}^{ab} \tilde{ \nabla}_a  \tilde{ \nabla}_b
\tilde{\phi} -\frac{d\tilde{V}}{d\tilde{\phi}} = 0 \,. 
\label{EframeKG}
\ee
If we know a solution of the Einstein equations with a 
minimally coupled scalar field as the matter source, it is 
possible 
to regard it as the Einstein frame representation of a 
scalar-tensor solution and to map it back to the Jordan 
frame representation. In general, the scalar field 
potential thus obtained in the Jordan frame is not 
motivated by a physical theory and the corresponding 
spacetime does not carry much physical meaning. This is the 
reason why this solution-generating technique has seen only 
limited applications in cases where the scalar field 
potential is absent \cite{NMCrefs, CMB}. However, in the 
particular application to the Fonarev spacetime 
\cite{Fonarev, MaedaFonarev} studied here, the Jordan frame 
potential $V(\phi)$ turns out to be physically well 
motivated.

$f({\cal R})$ theories of gravity \cite{reviews} are a 
subclass of scalar-tensor theories described by 
the action
\be
S = \int d^4 x \, \frac{\sqrt{-g}}{16\pi}  \, f({\cal R})   
\label{f(R)action}
\ee
{\em in vacuo}, where $f({\cal R})$ is a non-linear function of the Ricci 
scalar ${\cal R}$. By setting $\phi = f'({\cal R})$ and
\be \label{f(R)potential}
V(\phi)= \phi {\cal R}(\phi) -f\left( {\cal R}(\phi)  
\right) \,,
\ee
it can be shown that the action~(\ref{f(R)action}) is equivalent to the 
vacuum Brans-Dicke action \cite{reviews}
\be
S = \int d^4 x \, \frac{ \sqrt{-g}}{16\pi}  \left[ \phi 
{\cal 
R}-V(\phi) \right]  \,,
\ee
which has Brans-Dicke parameter  
$\omega=0$ and the potential~(\ref{f(R)potential}) for the Brans-Dicke scalar 
$\phi$.

The plan of this article is as follows. In Sec.~\ref{sec:2} 
we review the Fonarev solution of GR. In Sec.~\ref{sec:3} 
we obtain a new family of scalar-tensor solutions using the 
Fonarev spacetime as a seed. This family includes, as a 
special case, 
a solution previously reported in \cite{CMB} which is 
conformal to the Husain-Martinez-Nu\~nez geometry of GR 
\cite{HMN}. Another special case reproduces the 
Campanelli-Lousto solution \cite{CampanelliLousto}, which 
describes a wormhole \cite{VanzoZerbiniFaraoni}.  In 
Sec.~\ref{sec:4} 
we comment on the physical 
interpretation of the new family of solutions. 
Sec.~\ref{sec:5} explains how this new family is also a 
solution of a subclass of $f({\cal R})$ theories and 
Sec.~\ref{sec:6} contains a discussion and the conclusions.

\section{The Fonarev solution of general relativity} 
\label{sec:2} 
\setcounter{equation}{0}

The Fonarev solution of the Einstein equations 
\cite{Fonarev} is a spherically symmetric, dynamical, 
inhomogeneous, and asymptotically FLRW geometry sourced by 
a minimally coupled scalar field $\phi$ self-interacting 
with an exponential potential. The line element is
\begin{eqnarray}
d\tilde{s}^2 &=& -\mbox{e}^{8\alpha^2 at} \left( 
1-\frac{2m}{r} \right)^{\delta} dt^2 \nonumber\\
&&\nonumber\\  
&\, & +\mbox{e}^{2 at} \left[ \frac{dr^2}{ 
\left(1-\frac{2m}{r} \right)^{\delta} }
+ \left(1-\frac{2m}{r} \right)^{1-\delta} r^2 
d\Omega_{(2)}^2 \right] \nonumber\\
&&  \label{1}
\end{eqnarray}
and the matter scalar field is 
\be
\tilde{\phi}(t,r)=\frac{1}{ \sqrt{4\pi}} 
\left[ 2\alpha at +\frac{1}{2\sqrt{1+4\alpha^2}} \, \ln 
\left(1-\frac{2m}{r} \right) \right]\,,
\label{2}
\ee
with the scalar field potential 
\be \label{3}
\tilde{V}(\tilde{\phi}) = \tilde{V}_0 \, 
\mbox{e}^{-k\tilde{\phi}} \,.
\ee
This is a 3-parameter $\left( m,\alpha, a \right)$ family 
of solutions of the Einstein equations, where
\begin{eqnarray}
k &=& 8 \sqrt{\pi } \, \alpha \,, \label{4}\\
&&\nonumber\\
\delta &=& \frac{2\alpha}{\sqrt{1+4\alpha^2}} <1 \,, 
\label{5}\\
&&\nonumber\\
\tilde{V}_0 &=& \frac{a^2\left( 3-4\alpha^2\right)}{8\pi } 
\,.\label{6}
\end{eqnarray}
In order to guarantee a non-negative energy density for 
the scalar field it must be $\tilde{V}_0 \geq 0$, which 
implies that $|\alpha |\leq \sqrt{3}/2$.

This solution was introduced in \cite{Fonarev} and studied 
in~\cite{MaedaFonarev} and \cite{KastorTraschen2017}. 
Five-dimensional Fonarev solutions were given in 
\cite{Feinsteinetal2001} and it was shown recently that 
Fonarev solutions can be generated via dimensional 
reduction from Fisher-like brane solutions in $4+n$ 
dimensions \cite{KastorTraschen2017}.

The Fonarev line element is conformal to the 
Fisher-Buchdahl-Janis-Newman-Winicour-Wyman scalar field 
solution of the Einstein equations \cite{Fisher} (hereafter 
referred to simply as ``Fisher solution'')
\be
ds^2 = - {A(r)}^{\nu} dt^2 + {A(r)}^{-\nu} dr^2 +
{A(r)}^{1-\nu} r^2 d\Omega_{(2)}^2 \,, \label{Fisher}
\ee
\be 
\phi(r) = \phi_0 \ln A(r) \,,
\ee
where $\nu$ and $\phi_0$ are constants and $A(r) \equiv 
1-2m/r$. An interesting feature of the Fisher solution 
(pointed out in \cite{KastorTraschen2017}) is that the 
redshift factor for light travelling radially outward from 
a radius approaching the singularity diverges and these 
solutions have the properties of ``frozen stars'' like the 
Schwarzschild spacetime. The Fisher solution is the neutral 
limit of charged dilaton black holes 
\cite{chargeddilatonBH} and, therefore, the Fonarev 
solutions can be seen as limits of a family of dilaton 
black holes embedded in FLRW space 
\cite{KastorTraschen2017}.

For the special parameter values $\alpha=\pm \sqrt{3}/2$ 
the potential $\tilde{V}(\tilde{\phi}) $ vanishes 
identically and the Fonarev solution reduces to the 
Husain-Martinez-Nu\~nez solution of the Einstein equations 
\cite{HMN}, which is already known to be conformal to the 
Fisher solution \cite{HMN}. For $\alpha\neq \pm \sqrt{3}/2  
$ and  $\alpha a\neq 0 $, consider the new time coordinate 
$\tau$ defined by
\be 
\tau = \frac{ \mbox{e}^{4\alpha^2 a t}}{ 4\alpha^2 a}  \,,
\ee
which turns the Fonarev line element and scalar 
field~(\ref{1}) and (\ref{2}) into
\begin{eqnarray}
d\tilde{s}^2 &=& -{A(r)}^{\delta} d\tau^2 + \left( 
4\alpha^2 a \tau \right)^{1/(2\alpha^2)} 
\left[{A(r)}^{-\delta} dr^2 \right.\nonumber\\
&&\nonumber\\
&\, & \left. + 
{A(r)}^{1-\delta} r^2 d\Omega_{(2)}^2 \right] \,,
\end{eqnarray}
\be
\tilde{\phi} (\tau , r) = \frac{1}{4\sqrt{\pi }} \, \ln 
\left[ \left( 4\alpha^2 a \tau \right)^{1/\alpha} 
{A(r)}^{ \frac{1}{\sqrt{1+4\alpha^2}} } 
\right] \,.
\ee
The further time redefinition
\be
\eta = \frac{ \left( a\tau \right)^{1-\frac{1}{4\alpha^2}  
} }{ \left( 4\alpha^2 \right)^{\frac{1}{4\alpha^2}-1} 
\left( 
4\alpha^2 -1\right)a}  \label{TS}
\ee
for $\alpha^2\neq 0, 1/4$ and $a\neq 0$  turns the 
geometry 
and scalar field 
into
\begin{eqnarray}
d\tilde{s}^2 &=& \left( a\eta \right)^{ 
\frac{2}{4\alpha^2-1}} 
\left[  -{A(r)}^{\delta} 
d\eta^2 + {A(r)}^{-\delta} dr^2 \right. \nonumber\\
&&\nonumber\\
&\,& \left. + {A(r)}^{1-\delta} r^2 d\Omega_{(2)}^2 \right] \,,
\label{2.12}
\end{eqnarray}
\be
\tilde{\phi} \left(\eta , r\right) = 
\frac{1}{4\sqrt{\pi}} \, 
\ln \left\{ \left[ \left(4\alpha^2-1\right)a\eta\right]^{
\frac{4\alpha}{4\alpha^2-1}} {A(r)}^{ 
\frac{1}{\sqrt{1+4\alpha^2}} } \right\} \,.
\ee
The line element~(\ref{2.12}) is explicitly conformal to the 
Fisher one.

For $\alpha^2=1/4$ the coordinate 
transformation~(\ref{TS}) ceases to be valid but the 
line element reduces to
\be
d\tilde{s}^2= \mbox{e}^{ 2at} \left[  -{A(r)}^{\delta} dt^2 
+ {A(r)}^{-\delta} dr^2 + {A(r)}^{1-\delta} r^2 d\Omega_{(2)}^2 \right] \,,
\ee
which also is conformal to the Fisher line element~(\ref{Fisher}).

Special cases of the Fonarev solution include the following.

\subsection{Vanishing mass parameter}

When the mass parameter $m$ vanishes, the Fonarev solution 
reduces to the spatially flat FLRW metric 
\be
d\tilde{s}^2 = -\mbox{e}^{8\alpha^2 a t} dt^2 
+\mbox{e}^{2a t}\left( dr^2 +r^2 d\Omega_{(2)}^2 \right) 
 \label{7} 
\ee
and the scalar field is linear,
\be
\tilde{\phi}(t) = \frac{\alpha a t}{\sqrt{\pi }} \,.  
\label{8}
\ee
The redefinition of the time coordinate $d\tau = 
\mbox{e}^{4\alpha^2 a t} dt$ then recasts the line element as 
\be
d\tilde{s}^2 = - d\tau^2
+\left( 4\alpha^2 a \tau \right)^{\frac{1}{2\alpha^2}} 
\left( 
dr^2 +r^2 d\Omega_{(2))}^2 \right)  \label{10}
\ee
with scale factor $S(\tau)=\left( 4\alpha^2 a 
\tau \right)^{\frac{1}{4\alpha^2}}$ and matter scalar field 
\be
\tilde{\phi}(\tau)= \frac{\alpha}{\sqrt{ \pi }} \, \ln 
S(\tau) \,.
\ee
This universe is accelerated if $0< |\alpha| < 1$.

\subsection{Vanishing $\boldsymbol{a}$ parameter}

When the parameter $a$, which has the dimensions of an 
inverse time, vanishes the line element and scalar field 
reduce to
\be
d\tilde{s}^2 = - {A(r)}^{\delta} dt^2
+ {A(r)}^{-\delta}dr^2 
+ {A(r)}^{1-\delta}r^2 
d\Omega_{(2)}^2  \,, \label{11}
\ee
\be
\tilde{\phi}(t,r)= \frac{1}{4\sqrt{ \pi \left( 
1+4\alpha^2 \right)}} \, 
\ln A(r) \,. \label{12}
\ee
and the scalar field potential vanishes, $\tilde{V} \equiv 
0$. This is a static Fisher 
solution~(\ref{Fisher}) with $\nu=\delta$ and 
with areal radius $R(r)={A(r)}^{ 
\frac{1-\delta}{2}} r$. As is well known \cite{Fisher}, it 
exhibits a central singularity at $R=0$ (or $r=2m$).

\subsection{Parameter $\mathbf{ \boldsymbol{\alpha}^2=3/4 
}$}

The parameter $\alpha$ is related to the slope of 
the scalar field potential~(\ref{3}). In the special cases 
$\alpha=\pm \sqrt{3}/2$, the potential 
$\tilde{V}(\tilde{\phi})$ vanishes identically and the line 
element and scalar field reduce to
\begin{eqnarray}
d\tilde{s}^2 &=& - \mbox{e}^{6at} {A(r)}^{\pm \sqrt{3}/2} 
dt^2
+\mbox{e}^{2at} \left[ {A(r)}^{\mp \sqrt{3}/2}dr^2 \right. 
\nonumber\\
&&\nonumber\\
&\, & \left. + 
{A(r)}^{1\mp \sqrt{3}/2} r^2
d\Omega_{(2)}^2 \right] \,, \label{13}
\end{eqnarray}
\be
\tilde{\phi}(t,r)= \frac{1}{2\sqrt{ \pi }} \,
\left[  \pm \sqrt{3} \, at +\frac{1}{4} \ln A(r) \right]\,. 
\label{14}
\ee
This is recognized as the Husain-Martinez-Nu\~nez solution 
of GR \cite{HMN}. It is not obtained as the time 
development of regular Cauchy data because it contains a 
naked singularity part of the time and it exhibits an 
interesting phenomenology of apparent horizons which appear 
and disappear in pairs \cite{HMN}, and it constitutes one 
of two paradigmatic situations identified with 
time-evolving apparent horizons (the other situation is 
exemplified by McVittie-type solutions) \cite{mylastbook}. 
The special subcase $a=0$ eliminates the time dependence 
and reduces the Husain-Martinez-Nu\~nez solution to the 
Fisher spacetime. Assuming $a\neq 0$, the use of the time 
coordinate
\be
\tau(t) = \frac{ \mbox{e}^{3at}}{3a}
\ee
reduces the line element and scalar field to 
\begin{eqnarray}
d\tilde{s}^2 &=&  - {A(r)}^{\delta}d\tau^2
+ \left( 3a\tau \right)^{2/3} \left[ 
{A(r)}^{-\delta} dr^2 \right. \nonumber\\
&&\nonumber\\
&\, & \left. + {A(r)}^{1-\delta} r^2 d\Omega_{(2)}^2 \right] \,, 
\label{15}
\end{eqnarray}
\be
\tilde{\phi}(\tau,r)= \frac{1}{8\sqrt{ \pi }} \,
\ln \left[ A(r) \left(3a\tau \right)^{ \pm 4/ \sqrt{3} } 
\right]\,.
\label{16}
\ee
The geometry is asymptotically FLRW as $r\rightarrow 
+\infty$. The scale factor $S(\tau) \sim \tau^{1/3}$ 
corresponds to the stiff equation of state $P=\rho$ for the 
cosmic fluid equivalent of the free scalar field.

\subsection{Parameter $\mathbf{\boldsymbol{\alpha}=0}$}

This special case implies $k=\delta=0$ and the scalar field 
potential $\tilde{V}=\tilde{V}_0$ reduces to an effective 
cosmological constant $\Lambda= 8\pi \tilde{V}_0=3a^2$ with
\be
d\tilde{s}^2=  - dt^2
+ \mbox{e}^{2at} \left[
dr^2 + A(r) r^2 d\Omega_{(2)}^2 \right] \,,
\label{17}
\ee
\be
\tilde{\phi}(t,r)= \frac{1}{4\sqrt{ \pi }} \,
\ln  A(r) \,,
\label{18}
\ee
which can be seen as a time-dependent generalization of a 
Fonarev solution with $\nu=0$ (to which it reduces if  
$a=0$), which is asymptotically de Sitter. Because of 
formal 
similarities, the Husain-Martinez-Nu\~nez solution and its 
Fonarev generalization could be superficially seen as 
time-dependent 
generalizations of the Fisher solution but they are 
qualitatively different in the parameter range in which 
apparent horizons exist (the Fisher solution, by contrast, 
has no apparent/trapping horizons to cover the central 
naked singularity).

\section{Generating new Brans-Dicke solutions}
\label{sec:3}
\setcounter{equation}{0}

Following the method outlined in Sec.~\ref{sec:1}, assume 
now that the Fonarev solution $ \tilde{g}_{ab}$ of GR with matter 
scalar field $\tilde{\phi}$ is 
formally the Einstein frame representation of a solution of 
Brans-Dicke theory in the Jordan frame $\left( g_{ab}, \phi 
\right)$, where the scalar field is now the gravitational 
Brans-Dicke field, which is related to the Fonarev geometry 
by 
\begin{eqnarray}
\tilde{g}_{ab} &=& \phi \, g_{ab} \,,\label{19}\\
&&\nonumber\\
\tilde{\phi} &=& \sqrt{ \frac{ |2\omega+3|}{16\pi } } \, 
\ln \left( \frac{\phi}{\phi_0} \right) \,, \label{20}
\end{eqnarray}
where $\phi_0$ is a constant and $\omega\neq -3/2$. By 
inverting Eq.~(\ref{20}) one obtains
\be
\phi (t,r)= \phi_0 \, \mbox{e}^{ \frac{4\alpha at}{\sqrt{ 
|2\omega+3|}} }\left(1-\frac{2m}{r} \right)^{ \frac{1}{ 
\sqrt{|2\omega+3|(1+4\alpha^2)}} }  \label{21}
\ee
and the corresponding scalar field potential $V(\phi)$ is 
obtained from Eq.~(\ref{ppotential}) as
\be
V(\phi)=V_0 \phi^{2\beta} \label{22}
\ee
with
\be
\beta=1-\alpha \sqrt{|2\omega+3|} \,, \;\;\;\;\;\;
V_0=\tilde{V}_0  \phi_0^{2\alpha \sqrt{|2\omega+3|}} 
\label{23} \,.
\ee
As already remarked, in general the potential~(\ref{ppotential}) 
 generated by using the conformal 
transformation to the Jordan frame and a known GR solution  
as the seed is not of a form 
motivated by scalar field theories in particle physics or 
in cosmology. However, the 
power-law potential obtained in the Jordan frame from the Fonarev solution has been the 
subject of intensive studies in both cosmology and particle physics. It includes as 
special cases the mass potential $m_{\phi}^2 \phi^2/2$, the quartic 
potential $\lambda \phi^4$, 
and many quintessence potentials \cite{AmendolaTsujikawabook}.

The relation $g_{ab}=\phi^{-1} \tilde{g}_{ab} $ gives 
the Jordan frame line element
\begin{eqnarray}
ds^2 &=& - {A(r)}^{\frac{1}{\sqrt{1+4\alpha^2}} \left( 
2\alpha  -\frac{1}{\sqrt{ |2\omega+3|}} \right) } 
\mbox{e}^{ 4\alpha 
at \left( 2\alpha -\frac{1}{\sqrt{|2\omega+3|}} \right)} 
dt^2 \nonumber\\
&&\nonumber\\
&\, & +\mbox{e}^{ 
2at \left(1-\frac{2\alpha}{\sqrt{|2\omega+3|} } \right)}
\left[ {A(r)}^{ -\frac{1}{\sqrt{1+4\alpha^2} } 
\left( 2\alpha +\frac{1}{\sqrt{|2\omega+3|} } \right)}  
dr^2 \right. \nonumber\\
&&\nonumber\\
&\, & \left. + {A(r)}^{1-\frac{1}{ \sqrt{1+4\alpha^2}} 
\left( 
2\alpha + \frac{1}{\sqrt{|2\omega+3|} } \right) } r^2 
d\Omega_{(2)}^2 \right] \label{24}
\end{eqnarray}
(neglecting an irrelevant  overall multiplicative constant 
$\phi_0^{-1}$). We have a family of solutions of the vacuum 
Brans-Dicke field equations~(\ref{BDfe}), 
(\ref{box}) parametrized by the four 
parameters $\left( \omega, m, a, \alpha \right)$, of which 
$\omega$ is a parameter of the theory and the others are 
parameters of this specific family of solutions. By 
introducing the quantities \begin{eqnarray}
\gamma & \equiv & \frac{1}{ 
\sqrt{1+4\alpha^2}}\left(2\alpha + 
\frac{1}{\sqrt{|2\omega+3|} } \right) \,, \label{25}\\
&&\nonumber\\
\epsilon &\equiv& \frac{1}{
\sqrt{1+4\alpha^2}}\left(2\alpha -
\frac{1}{\sqrt{|2\omega+3|} } \right) \,, \label{26}
\end{eqnarray}
the new time coordinate
\be
\tau(t)= \frac{ \mbox{e}^{ 2\alpha a t \left( 2\alpha -
\frac{1}{\sqrt{|2\omega+3|} } \right)} }{2\alpha a 
\left( 2\alpha -
\frac{1}{\sqrt{|2\omega+3|} } \right)} \label{27}
\ee
(defined for $a\neq 0$ and $\alpha \neq 0, \frac{1}{2\sqrt{ 
|2\omega+3|}} $), and the FLRW scale factor
\be
S(\tau)= \left[ 2\alpha a \tau \left( 2\alpha -
\frac{1}{\sqrt{|2\omega+3|} } \right) \right]^{ \frac{ 
\sqrt{|2\omega+3|} -2\alpha}{ 2\alpha\left( 
2\alpha\sqrt{|2\omega+3|} -1 \right)}} \,, \label{28}
\ee
one can write the new family of solutions as  
\be
ds^2 =- {A(r)}^{\epsilon} d\tau^2
+S^2(\tau) \left[  {A(r)}^{ -\gamma}dr^2
+ {A(r)}^{1-\gamma } r^2 d\Omega_{(2)}^2 \right] 
\,,\label{29}
\ee
\be
\phi (\tau,r)= \phi_0 \left[ S(\tau) \right]^{ 
\frac{4\alpha }{\sqrt{|2\omega+3|} -2\alpha}}   
{A(r)}^{ \frac{1}{\sqrt{|2\omega+3|(1+4\alpha^2)} } } \,.  
\label{30}
\ee
When the mass parameter $m$ vanishes, the line element 
reduces to the spatially flat 
FLRW one
\be
ds^2=-d\tau^2 +S^2(\tau) \left( dr^2 +r^2 d\Omega_{(2)}^2 
\right) \,, \label{52}
\ee
while the Brans-Dicke scalar field is
\be
\phi(\tau)= \phi_0 \left[ S(\tau)\right]^{ \frac{4\alpha}{ 
\sqrt{ |2\omega+3|} -2\alpha} } \,.\label{53}
\ee
Therefore (when it is defined) the coordinate $\tau$ has 
the meaning of comoving time of the FLRW space in which the 
inhomogeneity is embedded.

By contracting the Brans-Dicke field equations~(\ref{BDfe}) and 
substituting $\Box \phi$ from Eq.~(\ref{box}) one obtains 
the Jordan frame Ricci scalar
\begin{eqnarray}
{\cal R} &=& \omega \, \nabla^c \ln \phi \nabla_c \ln 
\phi+\frac{1}{2\omega+3} \left( 3\, \frac{dV}{d\phi}+\frac{4\omega 
V}{\phi} \right) \nonumber\\
&&\nonumber\\ 
&=&  \frac{\omega}{\phi^2} \, \nabla^c\phi \nabla_c \phi 
+2V_0   \left[ \frac{3(\beta-1)}{2\omega+3} +1 \right]   
\phi^{2\beta-1}
\end{eqnarray}
and, using Eq.~(\ref{23}), 
\be
{\cal R} =   \omega \, \nabla^c \ln \phi \nabla_c \ln \phi 
+2V_0   \left[ 1- \frac{3\alpha \, \mbox{sign}\left( 
2\omega+3 \right)}{
\sqrt{|2\omega+3|}} \, \right]  \phi^{2\beta-1} \,.
\ee
Since 
\begin{eqnarray}
\nabla_{\mu} \ln \phi &=&  \frac{ 4\alpha a}{\sqrt{|2\omega+3|}} \, 
\delta_{\mu 0} \nonumber\\
&&\nonumber\\
&\, & + \frac{2m}{r^2 \left( 1-2m/r\right) 
\sqrt{ 
|2\omega+3|(1+4\alpha^2)}} \, \delta_{\mu 1} \,,\nonumber\\
&&
\end{eqnarray}
one obtains
\begin{eqnarray}
{\cal R} &=& -{A(r)}^{\frac{1}{ \sqrt{1+4\alpha^2}} \left(
\frac{1}{\sqrt{ |2\omega+3|} } -2\alpha\right)} \nonumber\\
&&\nonumber\\
&\, &
\cdot \frac{16\alpha^2 a^2 \omega \, \mbox{e}^{4\alpha at 
\left( \frac{1}{\sqrt{ |2\omega+3|}} 
-2\alpha\right)}}{|2\omega+3|} \, \nonumber\\
&&\nonumber\\
&\, & +\frac{4m^2\omega}{r^4 |2\omega+3|(1+4\alpha^2)} 
{A(r)}^{\frac{1}{ \sqrt{1+4\alpha^2}}\left( \frac{1}{\sqrt{ 
|2\omega+3|} } +2\alpha\right)-2} \nonumber\\
&&\nonumber\\
&\, & \cdot \,  \mbox{e}^{2at 
\left( 
\frac{2\alpha}{\sqrt{ |2\omega+3|}}-1\right)} \nonumber\\
&&\nonumber\\
&\, & +2V_0 \left[ 1-\frac{3\alpha}{\sqrt{ |2\omega+3|}} \, 
\mbox{sign}\left( 2\omega+3\right) \right] 
\phi_0^{2\beta-1} \nonumber\\
&&\nonumber\\
&\, & \cdot \, \mbox{e}^{ -4 \alpha at\left( 2\alpha 
-\frac{1}{\sqrt{|2\omega+3|}}\right)} 
{A(r)}^{\frac{1}{ 
\sqrt{1+4\alpha^2}}\left(\frac{1}{\sqrt{|2\omega+3|} } 
-2\alpha \right)}  \,. \nonumber\\
&& \label{ARicciscalar}
\end{eqnarray}
The three terms which add up to compose the Ricci scalar can vanish 
in special cases. The first term is absent if $\alpha=0$ or $a=0$ 
or $\omega=0$. The second term is absent if $m=0$ or $\omega=0$. 
The third term drops out if $V_0=0$ (which happens if $a=0$ 
or $\alpha^2=3/4$) or if $\phi_0=0$ (which is forbidden) 
or if $\alpha= \sqrt{|2\omega+3|} \, \mbox{sign}\left( 
2\omega+3 \right) /3 \equiv \alpha_{*}$.
Unless these three terms disappear simultaneously, there is 
a spacetime singularity at $r=2m$ for the parameter values 
for which the exponent of $A(r)$ in these terms is 
negative. 

Regardless of the possible presence of a 
spacetime singularity at $r=2m$, the scalar 
field~(\ref{21}) always vanishes, 
and the effective gravitational coupling $G_\text{eff}$ 
diverges, as $r\rightarrow 2m^{+}$, which is another 
physical pathology to be avoided ({\em 
e.g.}, \cite{Chase,SotiriouFaraoniPRL}). 

Assume that $\gamma \neq 1$: then the vanishing of the 
areal radius (given in eq.~(\ref{ArealRadius}) below) $R=0$ 
corresponds to 
$r=2m$ and to a central singularity if the Ricci scalar 
${\cal R}$ diverges there. 
When present, the singularity at $R=0$ is 
timelike. In fact, consider the surface 
of equation $\Psi(r)\equiv r-2m$. The direction of its 
normal is $ N_{\mu}= \nabla_{\mu}\Psi =\delta_{\mu r}$ and
\begin{eqnarray}
N_{c}N^c &=& g^{ab} N_a N_b= g^{rr} \nonumber\\
&&\nonumber\\
&=&
\mbox{e}^{2at\left(  \frac{2\alpha}{ 
\sqrt{|2\omega+3|}}-1\right)} {A(r)}^{\gamma} 
\end{eqnarray}
is positive for any $r>2m$, hence it is non-negative in the 
limit $r\rightarrow 2m^{+}$, making $N^c$  spacelike or 
null and the 
surfaces $\Psi(r)=$~const. timelike or null (further, if 
$\gamma=0$ then $N^cN_c$ is strictly positive and the 
singularity is timelike). The 
singularity of the  conformally related Fonarev solution is 
timelike \cite{MaedaFonarev} and the conformal map 
respects causality, hence  the 
singularity obtained as the limit $r\rightarrow 2m^{+}$ is 
timelike. Therefore, when there is a 
naked singularity, the geometry cannot be obtained as the 
development of regular data on an initial Cauchy 
surface.

\section{Interpretation of the solutions}
\label{sec:4}
\setcounter{equation}{0}

Let us interpret now the new solutions of Brans-Dicke 
theory found by mapping back the Fonarev solution of GR to 
the Jordan frame. We maintain the condition $|\alpha|\leq 
\sqrt{3}/2$ which now guarantees non-negativity of the 
Jordan frame potential $V(\phi)$. We look for possible 
apparent horizons and, if found, we 
attempt to identify them as black hole horizons or wormhole 
throats.\footnote{We identify a wormhole throat with an 
apparent 
horizon with radius corresponding to a double root of 
Eq.~(\ref{32}).  Other authors (such as  
\cite{HochbergVisser}) have more stringent definitions of 
wormhole throats.} In general, dynamical spacetimes do not 
admit event 
horizons and the best substitute for the notion of horizon 
is the apparent or trapping horizon (see Refs.~\cite{Booth, 
Alexreview, mylastbook} for definitions and reviews of the 
related literature). A naked singularity is one that is not 
covered by apparent black hole horizons. It has been shown 
by Agnese 
and La Camera that static spherical solutions of 
Brans-Dicke theory without a scalar field potential never 
describe black holes but can only represent wormholes or 
naked singularities \cite{AgneseLaCamera}. The possible 
extension of this result to the situations in which a 
scalar 
field potential is present and the spacetime is dynamical 
is worth exploring.

The areal radius which is read off the geometry~(\ref{29}) 
is
\be\label{31}
R \left( \tau, r \right)= S(\tau)  {A(r)}^{ 
\frac{1-\gamma}{2}} r \,.
\ee
Apart from the special parameter value $\gamma=1$, the 
coordinate radius $r=2m$ always  
corresponds to zero areal radius $R$. Therefore, for 
$\gamma \neq 1$ and assuming 
non-negative mass parameter $m$, the 
physical range of the radial coordinate is $r \geq 2m $ (or 
$R\geq 0$). The apparent horizons, when they exist, are 
located by the roots of the equation \cite{MisnerSharp}
\be
\nabla^cR \nabla_c R=0 \,. \label{32}
\ee
In coordinates $\left( t, r \right)$ the areal radius is 
\be
R(t,r)=\mbox{e}^{at \left( 
1-\frac{2\alpha}{\sqrt{|2\omega+3|}} \right)} {A(r)}^{ 
\frac{1}{2}-\frac{1}{ \sqrt{1+4\alpha^2}} \left( \alpha 
+\frac{1}{2\sqrt{|2\omega+3|}} \right)} r   
\label{ArealRadius}
\ee
and, in order for it to be well defined and positive, it 
must be $r>2m$ (except for the special case  
$\gamma=1$ in which the exponent of $A(r)$ vanishes). 
Equation~(\ref{32}) takes the form 
\begin{eqnarray}
a^2 \left( 1-\frac{2\alpha}{ \sqrt{|2\omega+3|}} \right)^2 
\, \mbox{e}^{2at\left(1-4\alpha^2\right)} r^2 {A(r)}^{ 
-\frac{4\alpha}{\sqrt{1+4\alpha^2}}+2 } \nonumber\\
\nonumber\\
=\left\{ 1-\frac{m}{r} \left[ 1+ 
\frac{1}{\sqrt{1+4\alpha^2}} \left( 
2\alpha+ \frac{1}{\sqrt{|2\omega+3|}} 
\right)\right]\right\}^2 \nonumber\\
\label{AHt}
\end{eqnarray}
in coordinates $\left( t, r \right)$. This form of 
Eq.~(\ref{32}) is useful when the time coordinate $\tau$ 
cannot be used. For the parameter values for which  
$\tau$ is well defined, Eq.~(\ref{32}) can be written in 
the form 
\be
{A(r)}^{ 2\left( 1-\frac{2\alpha}{ \sqrt{1+4\alpha^2} } 
\right)} r^2 S_{\tau}^2 = \left[ 1-\left( \gamma +1 \right) 
\frac{m}{r} \right]^2 \,.\label{33}
\ee
It is not possible to solve analytically Eq.~(\ref{AHt}) 
or~(\ref{33}) except for special points in parameter space. 
Likewise, their numerical solution requires the complete 
specification of the values of the parameters 
$\left(\omega, m, a,\alpha \right)$. Let us examine the 
solutions of Eqs.~(\ref{AHt}) and~(\ref{33}) in special 
cases.

\subsection{Special case~1 
($\mathbf{\boldsymbol{\alpha}=0}$)}

Let us consider the parameter value 
$\alpha=0$ which trivially satisfies the constraint 
$|\alpha|\leq \sqrt{3}/2$ and makes the coordinate 
transformation $t\rightarrow \tau$ invalid. In this 
case the scalar field potential $V=V_0 \phi^{2\beta}$ 
reduces to a mass term $m_{\phi}^2\phi^2/2$ with 
\be
m_{\phi}=\sqrt{2V_0}= \sqrt{2\tilde{V}_0}=\sqrt{ 
\frac{3a^2}{4\pi}} \,,
\ee
while    
$\gamma=|2\omega+3|^{-1/2} $. The Brans-Dicke spacetime is 
given by
\begin{eqnarray}
ds^2 &=&-  {A(r)}^{ -\, \frac{1}{ \sqrt{ |2\omega+3|}} }  
dt^2
+ \mbox{e}^{2at} \left[  
{A(r)}^{ -\, \frac{1}{ \sqrt{|2\omega+3|}}  } dr^2 \right.
\nonumber\\
&&\nonumber\\
&\, & \left. 
+ {A(r)}^{1 -  \frac{1}{ \sqrt{ |2\omega+3|}}  } r^2
d\Omega_{(2)}^2 \right] \,, \label{49}
\end{eqnarray}
\be
\phi (r)= \phi_0 
{A(r)}^{ \frac{1}{\sqrt{|2\omega+3|} } } \,.
\label{50}
\ee
Equation~(\ref{AHt}) becomes 
\be
a^2  \, \mbox{e}^{2at} r^2 {A(r)}^2 = \left[ 
1- \frac{m}{r} \left(  
1+\frac{1}{\sqrt{|2\omega+3|} } \right)\right]^2 \,,
\label{acci}
\ee
which cannot be solved analytically for general values of 
the parameters $\left( a,m,\omega \right)$. For 
illustration, we consider 
$a\neq 0$ in conjunction with the special values of the 
Brans-Dicke parameter $\omega=-2, 
-1$ for which $|2\omega+3|=1$. Then Eq.~(\ref{acci}) admits 
the single positive root
\be
r_\text{AH}= \frac{\mbox{e}^{-at}  }{a} 
\ee
or, since the areal radius is $R(t,r)=r\, \mbox{e}^{at}$,
\be
R_\text{AH}= \frac{1}{a} \,.
\ee
The ``background'' cosmology is obtained by letting $m$ go 
to zero and it is a de Sitter space with Hubble parameter 
$a$, constant scalar field $\phi_0$, and cosmological 
constant $\Lambda=8\pi V_0 \phi_0=3a^2\phi_0$. 
Therefore, this apparent horizon always coincides with 
the de Sitter (cosmological) horizon of the 
background, which is a null surface. 

The minimal physical requirement that the Brans-Dicke field
\be
\phi=\phi_0 \left(1-\frac{2m}{r} \right)
\ee
be positive imposes that $r>2m$. Then the apparent 
horizons exists only at comoving times
\be
t < a^{-1} \ln \left( \frac{1}{2ma} \right) \equiv t_{*} 
\,.
\ee
The effective 
coupling $G_{\text{eff}}$ neither diverges nor vanishes on 
this apparent horizon because $r=2m$ (where 
$A(r)$ vanishes) is distinct from $r_\text{AH}$, 
except at the time $t_{*}$. At $t=t_{*}$ it is 
$r_\text{AH}=2m$ 
and $\phi $ vanishes while $G_\text{eff}$ diverges.

\subsection{Special case~2 ($\mathbf{a=0}$)}

The time scale of variation of the new Brans-Dicke solution 
is, 
roughly speaking,  
$a^{-1}$, therefore the limit $a \rightarrow 0$ 
makes this time scale infinite, yielding a family of static 
solutions. 
In this case the coordinate 
transformation~(\ref{27}) degenerates and, using 
Eqs.~(\ref{24}) and~(\ref{21}), one obtains the geometry 
and Brans-Dicke field in coordinates $\left(t,r \right)$
\begin{eqnarray}
ds^2 &=&-  {A(r)}^{  \frac{1}{ \sqrt{1+4\alpha^2}} 
\left(2\alpha- \frac{1}{\sqrt{ |2\omega+3|} }\right)}  
dt^2 \nonumber\\
&&\nonumber\\
&\, & + {A(r)}^{ -\, \frac{1}{ \sqrt{1+4\alpha^2}} \left( 
2\alpha +\frac{1}{\sqrt{ |2\omega+3|} }\right)  } dr^2 
\nonumber\\
&&\nonumber\\
&\, & + {A(r)}^{1 -  \frac{1}{ \sqrt{1+4\alpha^2}} 
\left(2\alpha +\frac{1}{\sqrt{ |2\omega+3|} }\right)} 
r^2 d\Omega_{(2)}^2  \,, \label{a=0metric} \nonumber\\
\label{51}&&\\
\phi(r) &=& \phi_0 {A(r)}^{ \frac{1}{\sqrt{ |2\omega+3| 
(1+4\alpha^2)}} } \,,\label{a=0BDfield}
\end{eqnarray}
which are static, while $V(\phi)=0$. Equation~(\ref{AHt}) 
degenerates and admits the double root
\begin{eqnarray}
r_\text{AH}&=& m\left[ 1+\frac{1}{ \sqrt{ 
1+4\alpha^2}}\left( 
2\alpha +\frac{1}{\sqrt{|2\omega+3|}} \right) \right] 
\nonumber\\
&&\nonumber\\
&\equiv & \left(1+\gamma \right)m \,,\label{WAH}
\end{eqnarray}
which corresponds to a wormhole apparent horizon provided 
that $r_\text{AH}>2m$. This condition translates into
\begin{eqnarray}
\alpha &>&  \frac{\omega+1}{2\sqrt{2\omega+3}} \;\;\;\;\;\; 
\mbox{if} \;\;\; \omega>-3/2 \,, \label{cacc1}\\
&&\nonumber\\
\alpha &>&  \frac{-\left(\omega+2 
\right)}{2\sqrt{|2\omega+3|}} \;\;\;\;\;\; 
\mbox{if} \;\;\; \omega<-3/2 \,. \label{cacc2}
\end{eqnarray}
We further impose the condition $|\alpha|\leq \sqrt{3}/2$. 
Consider first the situation in which $\omega >-3/2$: then 
in order to satisfy both~(\ref{cacc1}) and $|\alpha|\leq 
\sqrt{3}/2$ it must be 
\be
\frac{\omega+1}{2\sqrt{2\omega+3}}< \frac{\sqrt{3}}{2} 
\,,\label{cacc3}
\ee
which is equivalent to $\omega+1 < \sqrt{3(2\omega+3)} $. 
If $\omega<-1$ this inequality is always satisfied while, 
if $\omega \geq -1$ both sides of~(\ref{cacc3}) are 
non-negative and we can square it, obtaining 
$\psi_1 (\omega) \equiv \omega^2-4\omega-8<0$. The 
parabola of equation $\psi_1 (\omega) $ has concavity 
facing upward, crosses the $\omega$-axis at $
\omega_{\pm}= 2\left(1\pm \sqrt{3} \right) $, 
and is negative if $\omega_{-} <\omega<\omega_{+}$. 
Therefore, the restriction $|\alpha|\leq \sqrt{3}/2$  
limits the range of the Brans-Dicke parameter to  
\be
 -\frac{3}{2} < \omega < 2\left(1+\sqrt{3} \right) \,.
\ee
Let us consider now the other situation $\omega<-3/2$: the 
restriction $|\alpha|\leq \sqrt{3}/2$ is compatible 
with~(\ref{cacc2}) only if
\be
-\, \frac{\left( \omega+2 \right)}{2\sqrt{2\omega+3}}< 
\frac{\sqrt{3}}{2} \,,\label{cacc4}
\ee
equivalent to $-\left( \omega+2\right) < 
\sqrt{3|2\omega+3|} $. If $ -2< \omega <-3/2$ the left hand 
side of~(\ref{cacc4}) is negative and its right hand side 
is non-negative, hence~(\ref{cacc4}) is always satisfied. 
If instead $\omega\leq -2$, then both sides 
of~(\ref{cacc4}) 
are non-negative and we can square this inequality, 
obtaining $ \psi_2(\omega) \equiv \omega^2 +10\omega 
+13<0$. The parabola $ \psi_2(\omega)$ has concavity facing  
upward, crosses the $\omega$-axis at $\omega_{\pm}= -5\pm 
2\sqrt{3}$, and is negative if $-5-2\sqrt{3} <\omega \leq 
-2$. Therefore the condition $|\alpha|\leq \sqrt{3}/2$ 
imposes the restriction on the range of the Brans-Dicke 
parameter 
\be
-5-2\sqrt{3}< \omega< -3/2 \,.
\ee

The wormhole apparent horizon has areal radius
\be
R_\text{AH}=m \left( 1+\gamma \right) 
\left( \frac{\gamma-1}{\gamma+1} 
\right)^{\frac{1-\gamma}{2}}  \,.
\ee
Let us discuss the causal nature of this apparent 
horizon, 
which is the surface of equation $f(r)=0$, where  $f(r) 
\equiv r-\left( \gamma+1 \right)m$. The normal to the 
surfaces $f=$~const. has components 
\be
N_{\mu}=\nabla_{\mu} f =\delta_{\mu 1} \,,
\ee
its norm squared is 
\be
N_aN^a =g^{ab} \nabla_a f \nabla_b f= g^{rr} = 
{A(r)}^{\gamma} \,,
\ee 
and on the apparent horizon it is  
\be
N_a N^a \Big|_{r_\text{AH}} =
\left( \frac{\gamma-1}{\gamma+1}  \right)^{\gamma}
\,.
\ee
If $\gamma>1$ the normal $N^a$ is spacelike and the 
apparent horizon is a 
timelike surface, while it is null if $\gamma=1$. The fact 
that this static apparent 
horizon is timelike for $\gamma>1$ is 
in apparent contradiction with the well known statement 
of GR \cite{HawkingEllis} that in stationary situations 
apparent horizons and event horizons (which are null) 
coincide. However 
there is no real contradiction here because the proof of 
this 
statement 
requires the dominant energy condition \cite{HawkingEllis}, 
which cannot be imposed on the Brans-Dicke scalar 
field.\footnote{The effective stress-energy tensor of 
$\phi$ 
in the right hand side of Eq.~(\ref{BDfe}) contains 
second order derivatives which have indefinite sign, 
contrary to the canonical products 
of first order derivatives. In addition to the fact 
that the sign of the coefficient $\omega$ can be negative, 
this fact makes the sign of the effective energy density 
indefinite.} The Brans-Dicke field $\phi$ does not diverge 
nor vanish on the apparent horizon~(\ref{WAH}).

The new family of static Brans-Dicke solutions 
obtained for $a=0$ contains, as a special case, another 
class of known solutions, the Campanelli-Lousto class 
\cite{CampanelliLousto}, which is obtained when 
\be
\omega>-\frac{3}{2} \,, \;\;\;\;\;
\alpha = \frac{1}{2\sqrt{2\omega+3}} \,.
\ee
The Campanelli-Lousto family is given by 
\begin{eqnarray}
ds^2 &=& -{A(r)}^{b_0+1}dt^2 +{A(r)}^{-a_0-1} dr^2 
+{A(r)}^{-a_0} r^2 d\Omega_{(2)}^2 \,,\nonumber\\
&&\\
\phi &=& \phi_0 {A(r)}^{\frac{a_0-b_0}{2}} \,,
\end{eqnarray}
where $a_0, b_0$, and $\phi_0$ are constants, with the 
first two related to the Brans-Dicke parameter by   
\be
\omega\left( a_0, b_0 \right)= \frac{-2\left( a_0^2+b_0^2 
-a_0b_0 +a_0 +b_0 \right)}{\left( a_0 -b_0 \right)^2} 
\label{omega(a0b0)}
\ee
and $V(\phi)=0$ \cite{CampanelliLousto}. The metric and 
Brans-Dicke field satisfy the field equations
\begin{eqnarray}
R_{ab}&=&\frac{\omega}{\phi^2} \nabla_a \phi \nabla_b \phi+
\frac{ \nabla_a \nabla_a \phi}{\phi} \,,\\
&&\nonumber\\
\Box \phi &=& 0 \,.
\end{eqnarray}
The correspondence with our $a=0$ 
line element~(\ref{a=0metric}) gives
\begin{eqnarray}
a_0 &=& -1+\frac{1}{\sqrt{1+4\alpha^2}} \left(  2\alpha
+\frac{1}{\sqrt{|2\omega+3|}} \right) \,,\label{a0}\\
&&\nonumber\\
b_0 &=& -1 - \frac{1}{\sqrt{1+4\alpha^2}} \left( 2\alpha 
-\frac{1}{\sqrt{|2\omega+3|}} \right) \,,\label{b0}
\end{eqnarray}
while the correspondence with our $a=0$ Brans-Dicke 
field~(\ref{a=0BDfield}) yields
\be
\frac{a_0-b_0}{2} = \frac{2\alpha}{\sqrt{1+4\alpha^2}} \,.
\ee
Setting, for consistency, this value equal to the value of 
$\left( a_0-b_0\right)/2$ obtained from Eqs.~(\ref{a0}) 
and~(\ref{b0}), which is $ 1/\sqrt{|2\omega+3|\left( 
1+4\alpha^2\right)} $, gives the special value of the 
$\alpha$-parameter
\be
\alpha= \frac{1}{2\sqrt{|2\omega+3|}} \,.
\ee
Then it must be
\begin{eqnarray}
a_0 &=& -1+\frac{2}{ \sqrt{1+|2\omega+3|}} 
\,,\label{a0again}\\
&&\nonumber\\
b_0 &=& -1 \,.\label{b0again}
\end{eqnarray}
Finally, the relation~(\ref{omega(a0b0)}) must also be 
reproduced. Using the values~(\ref{a0again}) and 
(\ref{b0again}) of $a_0$ and $b_0$, one has
\be
\frac{-2\left( a_0^2+b_0^2 
-a_0b_0 +a_0 +b_0 \right)}{\left( a_0 -b_0 \right)^2} 
=\frac{ |2\omega+3|-3}{2} \,.
\ee
If $\omega>-3/2$ this expression reduces\footnote{Since there is no potential, the restriction 
$|\alpha|\leq \sqrt{3}/2$ does not apply.} to $\omega$. 
Therefore, the Campanelli-Lousto family of static, 
spherically symmetric, asymptotically flat solutions of 
Brans-Dicke gravity is reproduced by our new solutions 
when $a=0$, 
$\omega>-3/2$, 
and $\alpha=\frac{1}{2\sqrt{2\omega+3}}$. It is now 
established that the Campanelli-Lousto spacetimes describe 
wormhole geometries \cite{VanzoZerbiniFaraoni}.

\subsection{Special case~3 
($\mathbf{\boldsymbol{\alpha}^2=3/4}$)}

In the special case  $\alpha^2=3/4$  it is 
$\alpha=\delta=\pm \sqrt{3}/2$ and
\begin{eqnarray}
ds^2 &=& - {A(r)}^{ \alpha -\frac{1}{2\sqrt{|2\omega+3|}} }
d\tau^2 \nonumber\\
&&\nonumber\\
&\, & +S^2(\tau) \left[ 
{A(r)}^{-\left(\alpha+\frac{1}{2\sqrt{ 
|2\omega+3|}}\right)} dr^2 \right.\nonumber\\
&&\nonumber\\
&\, & \left. + 
{A(r)}^{ 1-\left(\alpha+\frac{1}{2\sqrt{
|2\omega+3|}}\right)}r^2d\Omega_{(2)}^2\right] \,, 
\label{55}
\end{eqnarray}
\be
\phi( \tau, r) = \phi_0 \left[S(\tau)\right]^{  
\frac{4\alpha}{\sqrt{ |2\omega+3|}-2\alpha}} 
{A(r)}^{\frac{1}{2\sqrt{|2\omega+3|}} } \,. \label{56}
\ee
For $\alpha=\pm \sqrt{3}/2$ the Fonarev solution reduces to 
the Husain-Martinez-Nu\~nez geometry \cite{HMN}, as already 
noted. The Jordan frame counterpart of the 
Husain-Martinez-Nu\~nez solution of GR, 
which is a solution of Brans-Dicke theory without 
potential, was derived in Ref.~\cite{CMB} as 
\begin{eqnarray}
ds^2 &=& - {A(r)}^{ \alpha \left(1 -\frac{1}{\sqrt{ 
3(2\omega+3)} } \right)}d\tau^2 \nonumber\\
&&\nonumber\\
&\, & +\tau^{\frac{2\left( 
\sqrt{2\omega+3}-\sqrt{3}\right)}{
3\sqrt{2\omega+3}-\sqrt{3} }} \left[  
{A(r)}^{-\alpha \left(1+\frac{1}{\sqrt{3(2\omega+3)} } 
\right)} dr^2 \right.\nonumber\\
&&\nonumber\\
&\, & \left. +
{A(r)}^{ 1-\alpha\left(1+\frac{1}{\sqrt{
3(2\omega+3)}}\right)} r^2d\Omega_{(2)}^2\right] \,,
\label{57}\\
&&\nonumber\\
\phi( \tau, r) &=&  \tau^{  \frac{2}{\sqrt{ 3(2\omega+3)}-1}} 
{A(r)}^{\pm \frac{1}{2\sqrt{2\omega+3}} } \,. \label{58}
\end{eqnarray}
It is straightforward to check that, if $\omega>-3/2$ and 
$\alpha=+\sqrt{3}/2$, Eqs.~(\ref{57}) and (\ref{58}) 
coincide with our Eqs.~(\ref{55}) and (\ref{56}), 
respectively (our solution is slightly more general as it 
allows for $\omega<-3/2$ and $\alpha=-\sqrt{3}/2$). 
Therefore, for these parameter values, the Brans-Dicke 
solution which is the Jordan frame counterpart of the 
Fonarev solution of GR reduces to the 
Einstein frame sibling of the 
Husain-Martinez-Nu\~nez geometry found in \cite{CMB} and 
discussed in Refs.~\cite{Andrescousin, mylastbook}, in 
which it is 
found 
that only wormholes or naked singularities appear, as 
the remaining parameters $\omega$ and $a$ vary.

\subsection{The $\mathbf{ \boldsymbol{\omega} \rightarrow 
\infty }$ limit}

Let us analyze now the limit  $\omega 
\rightarrow \infty$, in which Brans-Dicke theory is usually 
believed to converge to GR \cite{Weinberg}. A complete 
and rigorous analysis of the limit of a family of solutions 
of a 
theory of gravity as a parameter of the family diverges 
would require 
coordinate-independent 
methods \cite{MacCallum}, but a 
more standard approach suffices here. Let us discuss 
the 
situation in which the parameters $\alpha$ and 
$\omega$ are independent of each other. Then, in the limit 
$\omega 
\rightarrow \infty$ the  scalar 
field~(\ref{21}) becomes constant, $\phi \rightarrow 
\phi_0$, and the potential reduces to a constant, $V(\phi) 
\rightarrow \tilde{V}_0 \phi_0^2$. This introduces the 
cosmological constant $\Lambda =8\pi \tilde{V}_0 \phi_0^2= 
a^2 \left(3-4\alpha^2\right) \phi_0^2$. The line 
element~(\ref{24}) reduces to the 
Fonarev line element~(\ref{1}) as $\omega \rightarrow 
\infty$. Therefore, one obtains the Fonarev geometry and a 
cosmological constant as the only effective matter source. 
This is not a solution of the vacuum Einstein equations (we 
know well that the Fonarev solution corresponds to a 
minimally coupled scalar field with an exponential 
potential and no cosmological constant as the matter 
source). The fact that vacuum or electrovacuum solutions of 
the Brans-Dicke field 
equations fail to reproduce the corresponding solution 
of GR is well known \cite{failure} and the reason for this 
behaviour has been discussed in the literature 
\cite{illusions}.

The failure of a Brans-Dicke solution to reproduce 
the corresponding GR solution as $\omega \rightarrow 
\infty$ has been linked to the fact that, in this limit, 
one expects the Brans-Dicke scalar field to have the 
asymptotics $ \phi=\phi_0 +\mbox{O}(1/\omega ) + \, ... $ 
while the solutions giving the ``incorrect'' 
limit exhibit instead the asymptotics $ \phi=\phi_0 
+\mbox{O}(1/\sqrt{|\omega |}) + \, ...$~ Therefore, while 
one would normally expect 
$\phi$ to become constant as $\omega \rightarrow \infty$ 
and all its gradients to disappear from the Brans-Dicke 
field equations~(\ref{BDfe}), when $\phi$ has the 
``anomalous'' behaviour the term 
\begin{eqnarray}
A_{ab} &\equiv &\frac{\omega}{\phi^2} \left( \nabla_a \phi 
\nabla_b 
\phi -\frac{1}{2} \, g_{ab} \nabla^c \phi \nabla_c \phi 
\right) \nonumber\\
&&\nonumber\\
&=&
\omega \left(\nabla_a \ln \phi \nabla_b \ln 
\phi -\frac{1}{2} \, g_{ab} \nabla^c \ln \phi \nabla_c 
\ln \phi \right) \nonumber\\
&& 
\end{eqnarray}
on the right hand side of these equations does not 
disappear but remains of order O($1$). This is exactly what 
happens with the conformal cousin of the Fonarev solution. 
In fact, Eq.~(\ref{21}) yields
\begin{eqnarray}
A_{\mu\nu} &=& \frac{16 \alpha^2 a^2 \omega \delta_{\mu 0} 
\delta_{\nu 0}  }{|2\omega+3|} 
\nonumber\\
&&\nonumber\\
&\, & + \frac{ 8m\alpha a \omega \left( 
\delta_{\mu 1} \delta_{\nu 0}+\delta_{\mu 0} \delta_{\nu 1} 
\right) }{|2\omega+3| \sqrt{ 
1+4\alpha^2} \, r^2 \left( 1-2m/r\right) }  \nonumber\\
&&\nonumber\\
&\, & + \frac{4m^2 \omega \delta_{\mu 1} \delta_{\nu 1}}{|2\omega+3|\left( 
1+4\alpha^2\right) r^4 \left( 1-2m/r\right)^2 } \,.
\end{eqnarray}
As $\omega\rightarrow \infty$ the tensor $A_{\mu\nu}$ tends 
to  
\begin{eqnarray}
A_{\mu\nu} &\approx & 8\alpha^2 a^2 \, \mbox{sign}\left( 
\omega \right) \delta_{\mu 0} \delta_{\nu 0} \nonumber\\
&&\nonumber\\
&\, & +\frac{ 4m\alpha a \, \mbox{sign}\left( 
\omega
\right)}{\sqrt{1+4\alpha^2} \, r^2 
\left(1-2m/r\right) }   
\left( \delta_{\mu 1} \delta_{\nu 0}+\delta_{\mu 0} 
\delta_{\nu 1}\right) \nonumber\\
&&\nonumber\\
&\, & +\frac{2m^2}{\left( 1+4\alpha^2\right) r^4 
\left( 1-2m/r\right)^2} \, \delta_{\mu 1} \delta_{\nu 1}\,.
\end{eqnarray}
which is of order unity.

In the special case~1 with $\omega=-2, -1$ and in the 
special case~2 previously examined,  the 
parameter $\omega$ can only 
assume values in a finite range and the limit $\omega 
\rightarrow \infty$ cannot be taken.

\section{Generating new solutions of $f({\cal R})$ gravity} 
\label{sec:5} \setcounter{equation}{0}

As is well known, $f({\cal R})$ gravity is equivalent to an 
$\omega=0$ Brans-Dicke theory with Brans-Dicke scalar 
$\phi=f'({\cal R}) $ subject to the  potential 
\be\label{potential}  
V(\phi)= {\cal R} f'({\cal R})-f({\cal R}) \,,
\ee 
where  ${\cal R}={\cal R}(\phi) $ is  a function of $\phi=f'({\cal 
R})$ usually
defined implicitly \cite{reviews}. 
One wonders whether the Jordan 
frame counterpart of the Fonarev solution can also be  
an analytic solution of $f({\cal R})$ gravity. For this to 
be true, one must set $\omega=0$ and (cf. Eq.~(\ref{23}))
\be
V_0 \left[ f'({\cal R})\right]^{2\beta}= {\cal R}f'({\cal 
R}) -f({\cal R}) \,, \;\;\;\;\;\;\;
\beta= 1-\alpha \sqrt{3} \,. \label{form}
\ee
It is easy to see that the functional form $f({\cal R})=\mu 
{\cal R}^n $ (where $\mu$ and $n$ are constants) 
satisfies Eq.~(\ref{form}) provided that
\be 
\beta=\frac{n}{2(n-1)} \,, \label{beta}
\ee
\be
V_0=\frac{n-1}{n^{2\beta}} \, \mu^{1-2\beta} 
\,,\label{Vquesta}
\ee
which require $n\neq 1$ (for $n=1 $ the theory reduces to GR). 

By comparing Eq.~(\ref{beta}) with $\beta= 1-\alpha 
\sqrt{3}$ it follows that the parameter $\alpha$ of 
the family of solutions is 
\be\label{alpha}
\alpha =  \frac{n-2}{2\sqrt{3} \left(n-1\right)} \,.
\ee
In particular, the conformal cousin of the 
Husain-Martinez-Nu\~nez solution obtained for $\alpha=\pm 
\sqrt{3}/2$ is a solution of $f( {\cal R})=\mu {\cal R}^n$ 
gravity for $n=1/2, 5/4$. For these values of the parameter $n$, 
${\cal R}^n$ gravity is ruled out by  Solar System 
experiments \cite{SolarSystem}, but it  
is anyway interesting to add one more formal  solution to the 
very scarce catalogue of analytic inhomogeneous solutions 
of $f({\cal R})$ gravity.

The potential~(\ref{potential}) 
is 
no longer required to satisfy $ |\alpha | \leq \sqrt{3}/2$, 
but one has $V_0>0$ if  
$n>1$.  Solar System constraints require $n=1+\sigma$ with 
$ \sigma =\left( -1.1\pm 1.2 \right) \cdot 10^{-5} $   
\cite{SolarSystem}, while any $f({\cal R})$ theory is 
required to satisfy $f'>0$ in order for the graviton to 
carry positive energy and $f''\geq 0$ for local stability 
\cite{reviews, mystabpaper}. In the cosmological 
setting these requirements are 
satisfied if $n=1+\sigma$ with $\sigma \geq 0$.   Then 
\be
\alpha=-\frac{ (1-\sigma)}{2\sqrt{3} \, \sigma} \,, 
\;\;\;\;\;\;\;\;\;
\beta=\frac{1+\sigma}{2\sigma}  \label{quella}
\ee
(with $\alpha<0$ for realistic theories), which gives the 
line element in the form 
\begin{eqnarray}
ds^2&=&-{A(r)}^{ -\, \frac{1}{\sqrt{ 1-2\sigma+4\sigma^2}}} 
\, \mbox{e}^{ \frac{2\left(1-\sigma\right) at}{\sqrt{3} \, 
\sigma} } dt^2 \nonumber\\
&&\nonumber\\
&\,&  + \mbox{e}^{\frac{2(1+2\sigma)at}{3\sigma}} \left[
{A(r)}^{\frac{1-2\sigma}{ \sqrt{1-2\sigma+4\sigma^2}}} 
dr^2 \right.\nonumber\\
&&\nonumber\\
&\, & \left. 
+ {A(r)}^{\frac{1-2\sigma}{ \sqrt{1-2\sigma+4\sigma^2}}-1 }
r^2 d\Omega_{(2)}^2 \right] \,.
\end{eqnarray}
For consistency it must then be
\be
\phi=f'({\cal R})=n\mu {\cal R}^{n-1}=\left( 1+\sigma 
\right)\mu {\cal R}^{\sigma} \,.
\ee
This equation can be checked using the 
expression~(\ref{21}) of the Jordan frame Brans-Dicke field 
obtained by setting $\omega=0$,
\begin{eqnarray}
\phi(t,r) &=& \phi_0 \, \mbox{e}^{\frac{4\alpha 
at}{\sqrt{3}} } {A(r)}^{ \frac{1}{\sqrt{3(1+4\alpha^2)}}} 
\\
&&\nonumber\\
&=& \phi_0 \, \mbox{e}^{\frac{-2(1-\sigma)at}{3\sigma} } 
{A(r)}^{ \frac{\sigma}{\sqrt{4\sigma^2 -2\sigma+1}} } \,,
\end{eqnarray}
where in the last equality we used Eq.~(\ref{quella})  and 
we note that $4\sigma^2 -2\sigma+1>0$ for any value of 
$\sigma$. By 
imposing that this scalar field be equal to $f'({\cal 
R})=\left( 
1+\sigma
\right)\mu {\cal R}^{\sigma} $ one obtains the 
expression of the Ricci scalar
\be
{\cal R} = \left[ \frac{\phi_0}{(1+\sigma)\mu} 
\right]^{1/\sigma} \, 
\mbox{e}^{\frac{-2(1-\sigma)at}{3\sigma^2} } 
{A(r)}^{ \frac{1}{ \sqrt{4\sigma^2 -2\sigma+1}} } 
\,,\label{minchia}
\ee
which can be compared with the 
expression~(\ref{ARicciscalar}) of the Ricci scalar 
already computed. For $\omega=0$ the latter reduces exactly 
to Eq.~(\ref{minchia}) upon use of Eqs.~(\ref{Vquesta}) and 
(\ref{quella}).

\section{Conclusions}
\label{sec:6}
\setcounter{equation}{0}

The Fonarev solution of the Einstein equations which has a 
scalar field with exponential potential as the matter 
source has been mapped to the Jordan frame of Brans-Dicke 
theory, generating a new 4-parameter family of solutions of 
the vacuum Brans-Dicke field equations ($\omega$ is 
a parameter of the theory and $\left( m, a, \alpha \right)$ 
are parameters of the specific solution of this family). 
Notably, the potential for 
the Brans-Dicke field in the Jordan frame is a power-law or 
inverse power-law potential, which is physically well 
motivated 
and is used extensively in  
cosmology and particle physics. The solutions are 
spherically symmetric, inhomogeneous, time-dependent, and 
asymptotically FLRW. Special cases include the conformal 
version of the Husain-Martinez-Nu\~nez solution of GR with 
free scalar field \cite{HMN} found in Ref.~\cite{CMB} using 
the same technique employed here, and the Campanelli-Lousto 
solution \cite{CampanelliLousto}, which is now known to 
describe a wormhole \cite{VanzoZerbiniFaraoni}, in 
agreement with our more general discussion of the case 
$a=0$.

It turns out that the conformal relative of the Fonarev 
geometry thus obtained is also a solution of  $f({\cal 
R})=\mu {\cal R}^n$ gravity. To the best of our 
knowledge 
only one other analytic solution of this theory 
with the same properties ({\em i.e.}, spherical, 
inhomogenous, dynamical, and asymptotically FLRW) is known 
\cite{Clifton}.

In order to interpret physically the conformal relative of 
the Fonarev solution it is necessary to solve the equation 
$\nabla^cR \nabla_c R=0$ locating the apparent horizons and 
assess when solutions exist. Unfortunately, this is a  
trascendental equation which would require the 
complete specification of the values of the four 
parameters and, even then, it cannot be solved 
analytically. We have, nevertheless, considered special 
cases for illustration, in which the geometry describes 
a wormhole throat or a naked  singularity.  
This result can be compared with the 
Agnese-La Camera theorem \cite{AgneseLaCamera} stating  
that the only static, spherical, and asymptotically flat 
solutions of 
the Brans-Dicke field equations without scalar field 
potential are geometries containing wormholes or naked 
singularities. 
(As a side note, all asymptotically flat or 
asymptotically de Sitter, spherical or cylindrical, black 
holes of Brans-Dicke theory 
with ``reasonable'' potentials are known and reduce to 
those of GR 
\cite{Hawking, SotiriouFaraoniPRL, BhattaRomano}.) It seems 
that the Agnese-La~Camera theorem might extend to 
the dynamical, asymptotically FLRW situations.  More general 
statements are considerably more complicated to establish  
than the proof  of \cite{AgneseLaCamera} because of the 
presence of the 
potential, the time dependence, and the asymptotics. We 
leave their investigation to future work.

\begin{acknowledgments}
It is a pleasure to thank Lorne Nelson for a discussion.
This work is supported by the Natural Sciences and 
Engineering Research Council of Canada.
\end{acknowledgments}


\end{document}